\documentclass[aps,twocolumn,english,superscriptaddress]{revtex4-1}
\usepackage[utf8]{inputenc}

\usepackage{amsmath}
\usepackage[encapsulated]{CJK} 
\usepackage[english]{babel}
\usepackage{graphicx}
\usepackage{hyperref,cleveref}
\usepackage{microtype}
\usepackage{siunitx}

\usepackage{texmf/abbreviations}
\usepackage{texmf/formulautils}

\hypersetup{
	colorlinks = true,
	linkcolor  = cyan,
	urlcolor   = blue,
	breaklinks = true
}

\crefname{equation}{Eq.}{Eqs.}
\Crefname{equation}{Eq.}{Eqs.}

\crefname{section}{Sec.}{Secs.}
\Crefname{section}{Sec.}{Secs.}

\crefname{figure}{Fig.}{Figs.}
\Crefname{figure}{Fig.}{Figs.}

\renewcommand{\paragraph}[1]{\noindent\textbf{#1}}

\begin{document}

\title{Quantum circuit optimization with deep reinforcement learning}

\begin{CJK*}{UTF8}{gbsn}
	\author{Thomas F\"osel}
	\affiliation{Max Planck Institute for the Science of Light, Staudtstr. 2, 91058 Erlangen, Germany}
	\affiliation{Physics Department, University of Erlangen-Nuremberg, Staudtstr. 5, 91058 Erlangen, Germany}
	\affiliation{Google Research, Mountain View, CA 94043, USA}
	
	\author{Murphy Yuezhen Niu}
	\affiliation{Google Research, Venice Beach, CA 90291, USA}
	
	\author{Florian Marquardt}
	\affiliation{Max Planck Institute for the Science of Light, Staudtstr. 2, 91058 Erlangen, Germany}
	\affiliation{Physics Department, University of Erlangen-Nuremberg, Staudtstr. 5, 91058 Erlangen, Germany}
	
	\author{Li Li (李力)}
	\affiliation{Google Research, Mountain View, CA 94043, USA}
	
	\begin{abstract}
		A central aspect for operating future quantum computers is quantum circuit optimization, \abbr{ie}, the search for efficient realizations of quantum algorithms given the device capabilities. In recent years, powerful approaches have been developed which focus on optimizing the high-level circuit structure. However, these approaches do not consider and thus cannot optimize for the hardware details of the quantum architecture, which is especially important for near-term devices. To address this point, we present an approach to quantum circuit optimization based on reinforcement learning. We demonstrate how an agent, realized by a deep convolutional neural network, can autonomously learn generic strategies to optimize arbitrary circuits on a specific architecture, where the optimization target can be chosen freely by the user. We demonstrate the feasibility of this approach by training agents on 12-qubit random circuits, where we find on average a depth reduction by $27\%$ and a gate count reduction by $15\%$. We examine the extrapolation to larger circuits than used for training, and envision how this approach can be utilized for near-term quantum devices.
	\end{abstract}
	
	\maketitle
\end{CJK*}

\section{Introduction}

The long-term goal of fault-tolerant large-scale quantum computing promises disruptive progress in multiple important areas of science and technology \cite{nielsen2002bible}, such as decryption \cite{shor1999factorization_algo}, database search \cite{grover1997search_algo}, quantum simulation \cite{lloyd1996universal_qu_simulators}, and for optimization problems \cite{farhi2000adiabatic_qc}. However, the considerable overhead required for fault-tolerant operations implies a strong incentive in the near term to explore applications that can already work on ``noisy intermediate-scale quantum'' (NISQ \cite{preskill2018nisq}) devices. With the help of suitable algorithms \cite{farhi2014qaoa,mcclean2016vqe}, such devices will be able to produce results that outperform classical computers, despite a moderately high rate of errors (``noise resilience''). Recently, a quantum advantage has been demonstrated experimentally for some first benchmark problems on NISQ devices \cite{arute2019google_qu_supremacy,zhong2020qu_advantage}.

Research on NISQ applications aims to deliver useful results while keeping the overall resources (number of qubits, computation time) required at a manageable level. Quantum circuit optimization (QCO) constitutes an essential step in addressing this challenge. Starting from a given quantum circuit, \abbr{ie}, a sequence of quantum operations (gates) acting on a set of qubits, the goal of QCO is to find a logically equivalent circuit that is likely to reduce the probability of errors, by having an overall shorter runtime and using fewer gates (possibly with emphasis on gates of a certain type).  This can be achieved mainly by compression (via combination or cancellation) and efficient parallelization of the quantum gates.

There has been important progress in QCO over the years, giving rise to a set of approaches like the family of T-par \cite{amy2014t_par,amy2019t_par}, TOpt \cite{heyfron2018t_opt} and T-Optimizer \cite{zhang2019t_optimizer}, and QCO based on ZX calculus \cite{kissinger2020py_zx}. These approaches are inspired by the requirements of large-scale fault-tolerant quantum computation and therefore work on scenarios which are quite removed from what will be encountered in NISQ hardware devices. In fact, they have been formulated in a hardware-independent way and rather operate on a global level.

What is especially needed for QCO on NISQ devices is, however, an approach that is able to take into account the optimization opportunities afforded by a more detailed consideration of the actual hardware (connectivity, gate set, \abbr{etc}) \cite{alexeev2021review_qu_software_stack}. Finding the right strategy for this task is a challenging problem. The benefits of any change in the quantum circuit are very context-dependent, especially if the circuit structure is irregular, and a greedy strategy (to always aim for the immediate improvement) is often not ideal.

Optimizing circuits by hand is very time-consuming, requires revising the strategy for every new platform, and can be prone to human errors. Hard-coded algorithmic strategies require considerable human implementation effort specific to the given architecture. In situations with complex decision-making, hard-coded strategies are also prone to unconsidered special cases impeding their ability to generate high-quality results. Optimization techniques like genetic algorithms or simulated annealing improve on these aspects, but lack computational efficiency because they do not select actions in a deliberate way, and they are unable to transfer the experience from earlier solved problem instances.

Optimizing strategies for complex decision-making problems is the goal of reinforcement learning (RL) \cite{sutton2018review_rl}, an important subfield of machine learning that goes beyond straightforward supervised learning. Based on repeated attempts, strategies which have been successful in the past are reinforced over time (exploitation), while alternatives are still occasionally tried out in a trial-and-error fashion (exploration). Due to the generality of this approach, RL is being employed in a broad and growing spectrum of applications, such as robotics \cite{kober2013rl_robotics}, video games \cite{mnih2013deepmind_atari,vinyals2019deepmind_starcraft}, board games like chess and Go \cite{silver2017deepmind_alphago_zero,silver2018deepmind_alpha_zero}, chemistry \cite{zhou2017rl_chemical_reactions,zhou2019rl_molecules,kearnes2019rl_molecular_graphs}, neural architecture search \cite{zoph2017rl_neural_arch_search}, and chip placement \cite{mirhoseini2020rl_chip_placement}. Because RL discovers strategies autonomously and therefore does not rely on a teacher, it often achieves super-human performance in situations where such a comparison can be made.

In recent years, RL has also been proposed for several problems in the field of quantum computing. Examples include quantum phase estimation \cite{palittapongarnpim2017rl_phase_estimation}, the design of quantum experiments \cite{melnikov2018rl_quantum_experiments}, quantum control \cite{bukov2018rl_quantum_ctrl,august2018rl_quantum_ctrl,niu2019rl_quantum_ctrl,porotti2019rl_quantum_ctrl}, quantum error correction \cite{foesel2018rl_for_qec,sweke2020rl_qec,andreasson2019rl_qec,nautrup2019rl_qec_codes} (alongside other machine learning approaches \cite{torlai2017ml_neural_decoder,baireuther2018ml_neural_decoder,krastanov2017ml_neural_decoder,johnson2017qvector}), and quantum metrology \cite{xu2019rl_qu_metrology,schuff2020rl_qu_metrology}.

In this work, we introduce deep reinforcement learning for quantum circuit optimization. Our approach enables the computer to autonomously discover strategies for reducing the depth and gate count of quantum circuits, for arbitrary gate sets and connectivity. It allows to choose the optimization target at will and permits extrapolation of the discovered strategy to larger circuits. Due to its flexibility and generality, the RL approach proposed here has the potential to become a valuable component of the toolbox needed to unlock the power of NISQ devices in the near future.

\begin{figure}
	\includegraphics{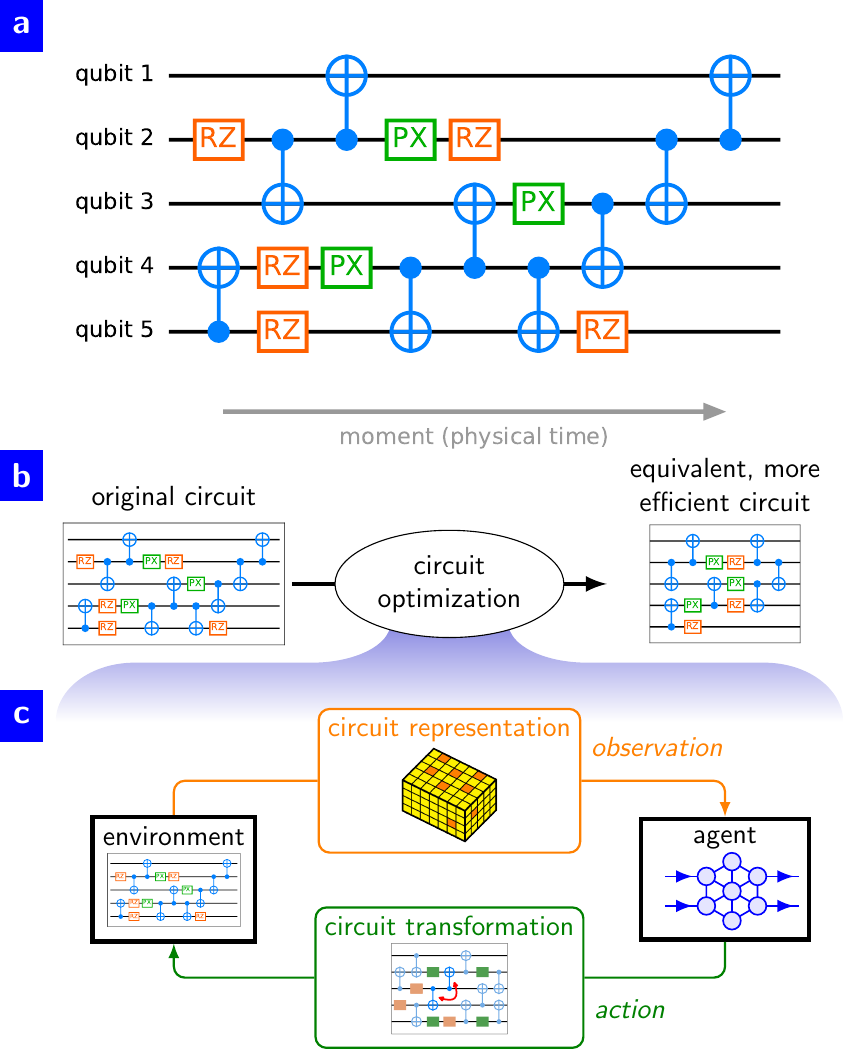}
	\caption{
		Overview.
		a) Diagram representation for quantum circuits. Each qubit is indicated with one line. The colored symbols represent operations (gates) on these qubits, with time increasing to the right.
		b) Quantum circuit optimization. For a given circuit, we aim to find a logically equivalent, but more efficient representation.
		c) Our reinforcement learning approach to quantum circuit optimization. Based on a diagram-like representation of the circuit, the agent, realized by a neural network, can choose between several circuit transformations to generate another, logically equivalent circuit; this process is repeated multiple times.
	}
\end{figure}

\section{Technique}
\label{sec:technique}

\subsection{Quantum circuit optimization as reinforcement learning problem}
\label{sec:technique:qco_as_rl_problem}

The goal of RL is to discover strategies for decision-making problems. This is described by an ``agent'' interacting with the rest of the world, the ``environment''. In several rounds, the agent receives information from the environment and, in response to this observation, chooses an action which alters the state of the environment. The agent is supposed to adapt its strategy so as to maximize a success measure, the ``reward''. More information is provided in \cref{sec:technique:reinforcement_learning}.

In the spirit of previous RL applications to quantum problems \cite{palittapongarnpim2017rl_phase_estimation,bukov2018rl_quantum_ctrl,august2018rl_quantum_ctrl,niu2019rl_quantum_ctrl,porotti2019rl_quantum_ctrl,foesel2018rl_for_qec,xu2019rl_qu_metrology,schuff2020rl_qu_metrology}, the obvious approach seems to let the agent build a circuit gate by gate to implement a certain target operation. However, this would come with two central problems here. First, it is extremely unlikely to find a suitable circuit by chance, so an untrained agent would in practice probably never see a positive reward signal. This problem is exacerbated by the fact that the gate set is typically not discrete, but gates can depend on continuous parameters. Second, in the particularly interesting quantum supremacy regime where the circuit cannot be simulated on a classical computer, there is the problem that even if one had found a valid circuit, verifying its correctness would be very hard and computationally expensive. Note that some tools like ZX calculus promise to arrive at a statement in polynomial time, but with the two possible results being positive or inconclusive whether two circuits are equivalent.

Therefore, we follow a different strategy that appears more promising: In QCO, it is common to start from a complete and correct, but typically inefficient circuit, and to progressively optimize it by applying a sequence of circuit transformations. However, it can be a formidable challenge to appropriately choose these transformations, and we make this decision the task of our agent. From the RL perspective, this means that \textit{the states are the circuits} and \textit{the actions are the circuit transformations}. By design, this approach immediately solves the challenge to finish with a correct, \abbr{ie}, logically equivalent, circuit: we can preserve this property for the full process by allowing in each step only equivalence transformations. In addition, our approach is also scalable, \abbr{ie}, it allows us to operate in the quantum supremacy regime: it is sufficient to verify equivalence for the few operations directly involved in an elementary circuit transformation, which is relatively cheap as long as all operations act only on a limited number of qubits.

Our general goal is to use RL to train a multi-purpose agent which afterwards will be able to optimize a wide class of circuits based on one given hardware architecture, without going through the RL procedure again in each instance. Sometimes, though, it can be helpful to fall back to the more restricted approach, where an agent is trained to optimize one specific circuit only.

For our approach, we need to carefully distinguish between different notions of time: the physical time which refers to the execution order of the operations in the circuit, and the agent time in which the actions, here to modify the circuit, are applied. In contrast to comparable previous works on RL for quantum problems \cite{palittapongarnpim2017rl_phase_estimation,bukov2018rl_quantum_ctrl,august2018rl_quantum_ctrl,niu2019rl_quantum_ctrl,porotti2019rl_quantum_ctrl,foesel2018rl_for_qec,xu2019rl_qu_metrology,schuff2020rl_qu_metrology}, these two time notions are here independent. From the agent's perspective, physical time is treated as an extra spatial dimension of the observation (\abbr{cmp} \cref{sec:technique:network_io}). In addition, there is the training time along which the weights and thus the behavior of the agent changes.

\subsection{Circuit transformations}
\label{sec:technique:circ_trafos}

Our goal is to optimize a circuit by applying a suitable sequence of circuit transformations. These transformations are based on local transformation rules. Such a rule could be to remove two subsequent operations that cancel each other; or, to reverse the order of two operations that commute (or anti-commute, as a global phase does not matter). Note that one given rule might be applicable to multiple locations in the circuit, yielding an independent transformation for each of them.

In our approach, we distinguish two kinds of transformation rules, which we refer to as ``hard'' and ``soft'' rules, and likewise call the transformations they induce hard and soft. Hard rules are those which are always advantageous, like the example of cancelling operations (if the goal is to shrink the circuit, \abbr{cmp} \cref{sec:technique:reward}). Soft rules, on the other hand, are those where the benefit of the transformation depends on the context. For example, for two commuting operations there might be a preferable order contingent on the surrounding operations, and dependent on the current configuration, it can make sense to exchange them or not. Also, it is not uncommon for soft transformations to be beneficial only in combination with several other hard or soft transformations.

In our approach, the decision of the agent is restricted to which soft transformation to apply next. This is always followed by a ``pruning'' step where automatically all permissable hard transformations are applied. In this sense, the decision of the agent can also be interpreted as not only selecting the soft, but implicitly also all ensuing hard transformations.

Currently, we determine and implement the small set of acceptable transformation rules by hand. The effort for this is manageable because only a moderate number of real-world quantum architectures exists, each with only a limited number of rules according to its specific gate set. As soon as we have arrived at a proper set of rules for one architecture, we can deal with any circuit on it.

\subsection{Learning algorithm}
\label{sec:technique:reinforcement_learning}

The full RL training process is divided into several episodes. In our case, an episodes corresponds to one optimization run for a specific circuit (which might change between episodes). Each episode itself is a sequence of $T$ subsequent steps, in which the agent changes the state $s_t$ with its action $a_t$ into the new state $s_{t+1}$. For this, the agent receives a reward $r_t$; in our case, the reward indicates the improvement of the circuit compared to the previous one. The objective in RL is to maximize the cumulative reward $\sum_{t=0}^{T-1}r_t$. This makes the return
\begin{equation}
	R_t = \sum_{t'=t}^{T-1} \gamma^{t'-t} r_{t'},
\end{equation}
where $\gamma$ is a discount rate, an important quantity: For $\gamma=1$, maximizing $\mathbb{E}\big[\sum_{t=0}^{T-1}r_t\big]$ would be equivalent to greedily maximizing $\mathbb{E}[R_t]$ in each time step; in practice, however, a value $\gamma<1$ often makes learning more stable and efficient.

The agent is typically implemented by a neural network, or in general by an arbitrary parameterized ansatz. The interpretation of its output is defined by the RL algorithm, which also specifies how to train the agent's parameters $\theta$. In this work, we use proximal policy optimization (PPO \cite{schulman2017ppo}), from the family of advantage actor-critic (AAC) methods. An AAC agent computes two quantities: the policy $\pi_\theta(a|s)$, according to which the actions are probabilistically chosen; and an estimator for the state value $V_\theta(s)$, that is the expectation value for the return $R$ when starting from state $s$ and following policy $\pi_\theta$. In our case, the meaning of the state value $V_\ast(s)$ for the best policy $\pi_\ast$ is the further potential for optimizations of the circuit representing state $s$.

AAC methods cooptimize policy $\pi_\theta$ and state value $V_\theta$ during training \cite{sutton2018review_rl}: Actions $a_t$ are reinforced according to their advantage $r_t+\gamma V_\theta(s_{t+1})-V_\theta(s_t)$, whose expectation value tells how much the average return $\mathbb{E}_{\pi_\theta}[R_t|s_t,a_t]=\mathbb{E}_{\pi_\theta}[r_t+\gamma V_\theta(s_{t+1})|s_t,a_t]$ for the chosen action $a_t$ exceeds the average over all actions, that is $V_\theta(s_t)$. Simultaneously, the state-value estimator $V_\theta(s)$ is refined (and adjusted to changes in the policy $\pi_\theta$) so as to satisfy the Bellman equation $V_\theta(s)=\mathbb{E}_{\pi_\theta}[r+\gamma V_\theta(s')|s]$. In the simplest case, this can be achieved by updating $\theta$ according to the learning gradient
\begin{equation}
	g =
	\mathbb{E}_{\pi_\theta}\bigg[
		\sum_{t=0}^{T-1} \big(r_t+\gamma V_\theta(s_{t+1})-V_\theta(s_t)\big) \frac{\partial}{\partial\theta}\Big[\ln\pi_\theta(a_t|s_t) - \lambda V_\theta(s_t)\Big]
	\bigg]
\end{equation}
where the coefficient $\lambda$ weights the cost term for the state value. PPO extends this basic AAC scheme in several ways for more powerful and efficient learning.

\subsection{Reward}
\label{sec:technique:reward}

The central motive behind QCO is often to decrease the probability for potential errors to occur. More generally speaking, we always want to improve some arbitrary desirable property of a circuit, which is defined by the user. The reward, which decides towards which behavior RL steers the agent, needs to be constructed according to this goal. Let $q(s)$ be a function which quantifies the desirable property for a given circuit $s$, and is to be minimized. A successful agent is expected to progressively improve the circuit over the course of an episode, so the goal is to minimize $q(s_T)$ at the final step $T$. We drive the agent towards this behavior with an immediate reward of the form
\begin{equation}
	r_t = -\big(q(s_{t+1}) - q(s_t)\big).
\end{equation}
This provides better time resolution than a final reward, such as $r_t\sim-q(s_T)\cdot\delta_{t,T-1}$ or $r_t\sim -(q(s_T)-q(s_0))\cdot\delta_{t,T-1}$, and therefore promises more efficient learning.

Now, we will derive a simple form for $q(s)$ which estimates the error probability for circuit $s$, or -- to be precise -- is a monotonously increasing function thereof. The quite simplistic, but still reasonable model behind it is to assume that each qubit loses information with a certain rate $\Gamma$ while being idle, and that every gate induces additional errors. Let $u_k$ indicate how much the $k$-th gate underperforms compared to decay with rate $\Gamma$ during idle times. Then, the probability to execute a circuit error-free is given by
\begin{equation}
	P_\mathrm{success} = \mconst{e}^{-m\Gamma T} \cdot \prod_{k=1}^{n} u_k
\end{equation}
where $m$ is the number of qubits, $n$ is the number of gates in the circuit, and $T$ denotes the total runtime. Our goal to maximize $P_\mathrm{success}$ is equivalent to minimizing $q = -\ln P_\mathrm{success} = m\Gamma T - \sum_{k=1}^{n} \ln u_k$.

During application of our RL approach, we will have to make concrete choices for the parameters, but we want to remain as general as possible. The least special assumption we can make is to attribute unit cost to all operations, \abbr{ie}, an equal execution time $\alpha/(m\Gamma)$ and an equal underperformance factor $u_k=\mconst{e}^{-\beta}$. In this case, $q$ can be expressed via the circuit depth $d$ and gate count $n$:
\begin{equation}
	q(s) = \alpha d(s) + \beta n(s).
	\label{eq:def_quality_fn}
\end{equation}
Explicitly, we will set $\alpha=1$ and $\beta=0.2$ in \cref{sec:results}. However, we emphasize that $q(s)$ in \cref{eq:def_quality_fn} can be replaced with any function of a circuit, and because we employ model-free RL, nothing in the learning algorithm would need to be changed.

\subsection{Representation of observations and actions}
\label{sec:technique:network_io}

\begin{figure*}
	\includegraphics{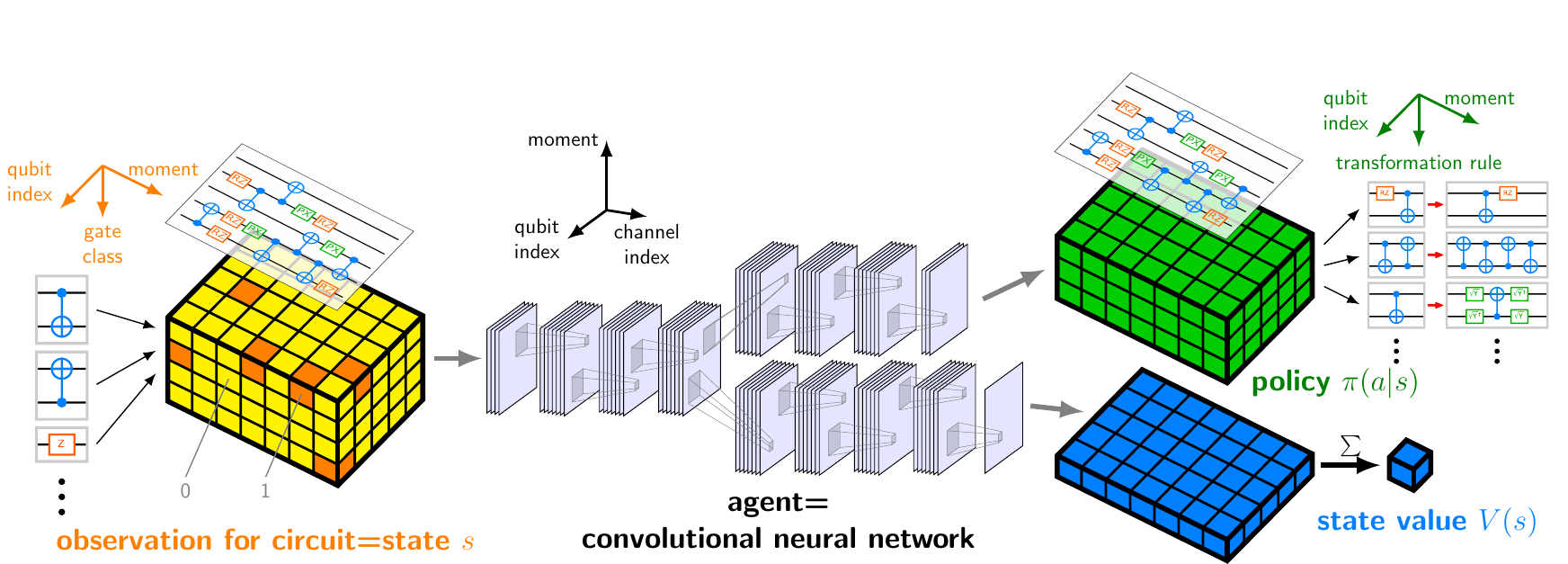}
	\caption{
		Deep convolutional network architecture of our RL agent. As observation, the agent receives a complete description of the state $s$, \abbr{ie}, the quantum circuit. The input neurons are arranged on a 3D grid, whose axes correspond to qubit index, moment and gate class. This information is processed through a stack of multiple convolutional layers, where qubit index and moment are treated as spatial dimensions and the gate classes as input color channels. For the output, the agent computes two quantities: (i) The policy $\pi(s|a)$, according to which the actions $a$ in state $s$ are probabilistically chosen. Every action, \abbr{ie}, circuit transformation, is mapped uniquely to one policy output neuron; the remaining neurons are disabled with an action mask. And (ii), the state value $V(s)$, which helps to update the policy $\pi(s|a)$ more efficiently during training. For us, $V(s)$ has the meaning of the optimization potential for the circuit.
	}
	\label{fig:network_arch}
\end{figure*}

With perfect information about the environment state, \abbr{ie}, the circuit, the agent can make the best decisions. Therefore, we provide a complete description of the circuit as observation to the agent. In quantum computing, circuit diagrams are widely used to visualize circuits for humans. Our format is inspired by them, but adjusted to the conditions of a neural network: We reserve one input neuron for each possible operation in the circuit, and activate neurons depending on whether the corresponding operation is actually present. These input neurons are arranged on a 3D grid, whose axes correspond to qubit index, moment (time step) and gate class. For gate types with a discrete number of instances, each of them forms its own gate class. For gate types which depend on (continuous) control parameters, we distinguish separate classes for special parameter values (\abbr{eg}, for Z rotations by $\pi/2$, $\pi$ and $3\pi/2$), and group all remaining instances of this gate type into one class. This allows to represent quantum circuits for arbitrary choices of their continuous parameters, but to distinguish gates for which additional transformation rules apply. To characterize multi-qubit gates, in principle all qubit indices would have to be indicated. In a 1D chain of qubits where gates are allowed only between nearest neighbors, as in our examples in \cref{sec:results}, we can characterize two-qubit gates simply by their lowest-index qubit, and specify by the gate class which roles the qubits play (\abbr{eg}, control \abbr{vs} target qubit for a CNOT gate).

A particular challenge in this project was to design the output format to represent the policy of the agent. The basic problem is that the actions, \abbr{ie}, the transformations of the circuit, are defined by the underlying transformation rule and all the gates affected by this transformation (which are characterized by their location in the circuit). Due to combinatorial explosion, the usual approach for discrete action spaces, to reserve one output neuron for every possible action, is not feasible here. One alternative would be to sort the transformations in a certain way and assign them consecutively to a queue of output neurons. However, this approach is not very promising either since the meaning for each output neuron would change for every circuit, potentially depending on aspects of very distant locations of the circuit, and therefore be very hard for the network to understand. Another alternative would be Q-learning where the agent, instead of computing the action value $Q(s,a)$ for state $s$ and action $a$, computes $Q(s,s')$ for input state $s$ and output state $s'$ \cite{zhou2019rl_molecules}; while conceptually this should work, this RL variant is less widely used and therefore not as well explored, and is restricted to Q-learning.

Instead, inspired by the representation of chess moves in \cite{silver2018deepmind_alpha_zero}, we follow a different approach where we associate transformations with certain locations in the circuit, in a way that uniquely characterizes which gates they affect. In our case, we use the following mapping: for a transformation involving a single gate, we indicate the moment and the smallest qubit index of the gate; for a transformation involving two gates, we indicate the smallest qubit index shared between both gates, and the moment of the first one. By providing multiple neurons per location to indicate the underlying rule, we can achieve an injective mapping from transformations to output neurons (for the policy). Therefore, also these neurons are arranged on a 3D grid, whose axes correspond to qubit index, moment and transformation rule. There can be neurons to which no transformation is associated; we disable them with an action mask, whose value changes with the input circuit.

Besides solving the problem to keep the total number of output neurons at a moderate level, another central advantage of this format is that we can exclusively use convolutional layers \cite{lecun1998lenet} to process the observation into the policy, treating qubit and moment as spatial dimensions, and the remaining grid axis (gate class and transformation rule, respectively) as input ``color'' channels. Also to compute the state value, we use convolutional layers (with one output channel), and eventually average over the spatial dimensions. \Cref{fig:network_arch} illustrates the architecture of our deep convolutional network. The weight sharing in the convolutional layers contributes to efficient and robust learning, and a \textit{fully} convolutional architecture will allow us to directly extrapolate to different circuit sizes (see \cref{sec:results:extrapolation_to_large}).

\subsection{RL problem classification}

The RL problem in this article can be classified as a Markov decision process (MDP) with perfect information (since the circuit representation, which is given to the agent as its input, completely describes the state of the environment). The set of all circuits comprises the state space. The set of possible circuit transformations represent the action space, which is therefore discrete, and its size depends on the state, \abbr{ie}, the circuit. The environment is deterministic: a fixed transformation (action) on a fixed circuit (state) always leads to the same outcome. Because the goal is to optimize a property which can be evaluated for any single given circuit, we can construct an immediate reward scheme (as opposed to situations where the reward is given only at the end of an episode).

\section{Results}
\label{sec:results}

In exploring the power of our RL approach, we need to select both a specific architecture (available gate set, processor layout, qubit connectivity) as well as the family of quantum circuits on whose optimization we want to focus.

\paragraph{Gate set} For our simulations, we consider the gate set consisting of Z-Rotation, Phased-X and Controlled-Not (CNOT) gates. Together, they form a universal gate set. Whereas  Z-Rotation and Phased-X are actually gate classes parameterized by 1 and 2 continuous variables, respectively, the CNOT is one fixed gate.

For our purposes, Z-Rotation, Phased-X and CNOT is a decent gate set because on the one hand, it induces a rich set of relatively simple transformation rules. On the other hand, they are quite similar to current real-world quantum hardware, such as Google's Bristlecone (Z-Rotation, Phased-X and Controlled-Z gate \cite{kelly2018bristlecone}) and Sycamore (Z-Rotation, Phased-X and fermionic simulation gate \cite{arute2019google_qu_supremacy,google2020hartree_fock,harrigan2021qaoa}) processor: The CNOT gate differs from the Controlled-Z gate only by local gates (\abbr{eg}, Hadamard or Sqrt-Y) on the target qubit, and the fermionic simulation gate is a generalization of the Controlled-Z gate.

\subsection{Training on random circuits}
\label{sec:results:rand_circs}

\begin{figure*}
	\includegraphics{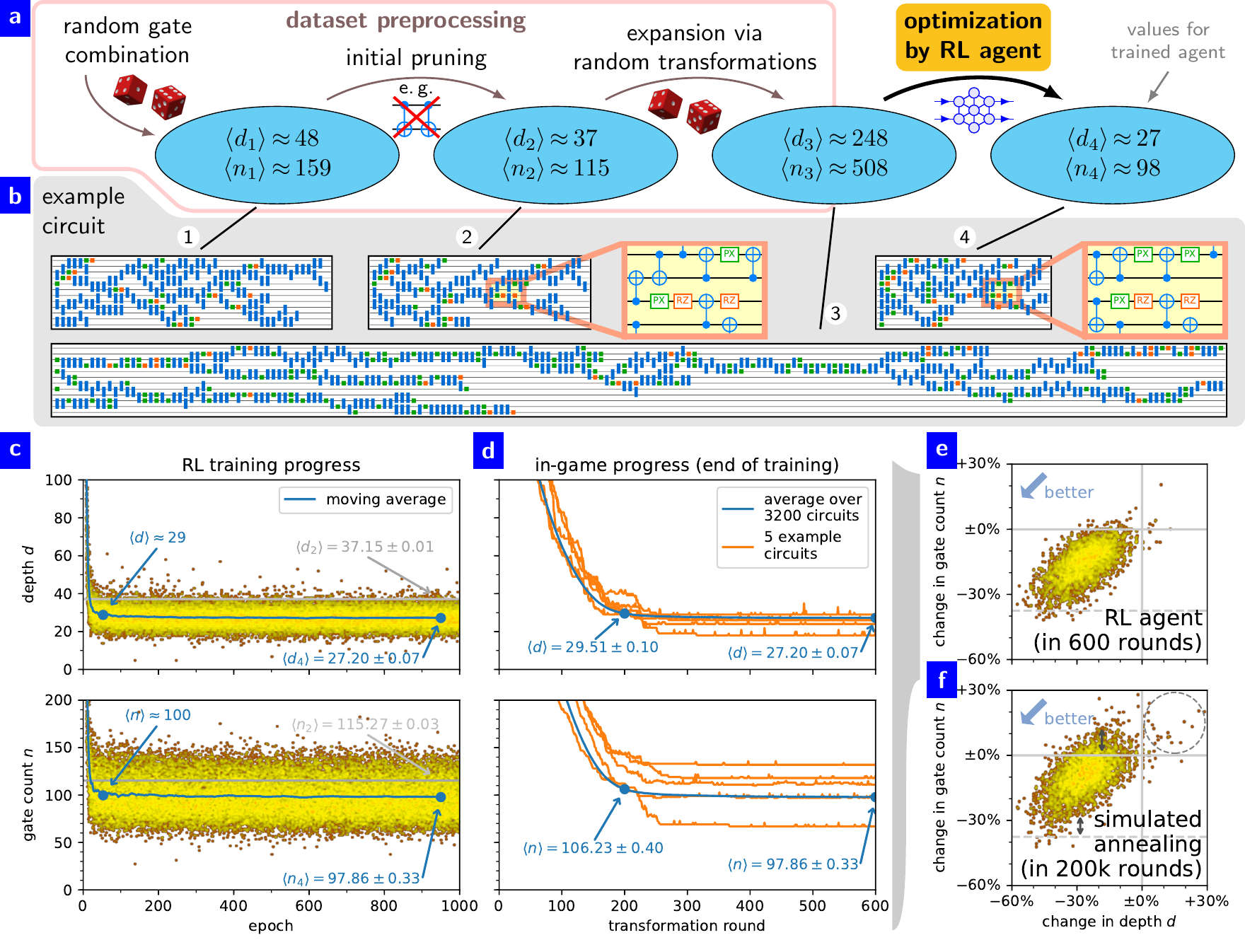}
	\caption{
		Training on random circuits.
		a) Circuit processing pipeline (see \cref{sec:results:rand_circs} for details). After choosing an initial circuit by randomly combining gates, a pruning step follows where all ``trivial'' optimizations are applied. Afterwards, $500$ random transformations are performed on this circuit, which turns out to significantly increase their depth $d$ and gate count $n$. These expanded circuits are then used as the starting point of the episodes to train and evaluate the RL agent.
		b) Diagrams illustrating the evolution of one example circuit through this pipeline.
		c) Learning progress during training, demonstrating how the agent improves in reducing both the depth $d$ (top) and the gate count $n$ (bottom) of the circuits. The point cloud indicates, for all episodes during training, the corresponding quantity in the final time step. The blue curve shows the moving average over the latest $10\%$ of epochs. For comparison, the gray line indicates the corresponding averages after pruning; already early in the training, the agent falls below this level for both quantities.
		d) In-game progress at the end of the learning process, showing for $5$ episodes during the last epoch (orange) how the agent progressively optimizes $(d,n)$ during an episode. The blue curve indicates the average over all episodes in the last $100$ epochs of training.
		e) Relative improvement achieved by the RL agent, in reference to the corresponding circuit size after pruning. Each point corresponds to one episode during the last $100$ epochs.
		f) Comparison with circuit optimization by simulated annealing (see \cref{sec:results:rand_circs} for details). The graphical depiction and the considered circuits are equivalent to (e), which makes them directly comparable.
	}
	\label{fig:results_small_rand_circs}
\end{figure*}

\paragraph{Dataset} While a quite large number of quantum circuits for various applications and architectures can be found in publications of the recent years, to the knowledge of the authors a joint dataset of such circuits does not exist, and collecting them by hand would have been very tedious. At the stage of developing the technique and verifying its feasibility, we therefore decided to employ randomly generated circuits. Although the majority of them is likely not to be useful for actual quantum applications, they have the big advantage that they can be easily generated in large numbers. Due to the irregular structure of such circuits, the benefits of transformations are strongly context-dependent, and therefore random circuits constitute a case where formulating a generally applicable strategy would be enormously complex.

We now briefly describe our method to generate these random circuits. We consider $12$ qubits arranged in a 1D line; two-qubit gates, here CNOTs, are only allowed between nearest neighbors. First, we build a sequence of $150$ gates whose type (either CNOT or local) and the qubits it acts on (constrained by the connectivity) are chosen randomly. Because local gates might be decomposed into a Z-Rotation and a Phased-X gate, we obtain a slightly higher value for the mean number of hardware gates, $\langle n_1\rangle\approx159$, and for the mean depth we find $\langle d_1\rangle\approx48$. For many of these circuits, some hard rules (\abbr{cmp} \cref{sec:technique:circ_trafos}) can immediately be applied, \abbr{ie}, there are some ``trivial'' optimizations; for example, two subsequent CNOTs on the same qubits directly cancel. To apply these transformations, which we refer to as pruning, reduces the mean depth to $\langle d_2\rangle=37.15\pm0.01$ and the mean gate count to $\langle n_2\rangle=115.27\pm0.03$. From here, we continue by applying random transformations (based on soft rules, each time followed by a pruning step), which turns out to make the circuit much larger and hence very inefficient. After $500$ of these transformations, we find a mean depth of $\langle d_3\rangle\approx248$ and a mean gate count of $\langle n_3\rangle\approx508$.

We use these randomly expanded circuits as the training tasks for the RL agent, where each of them defines the starting point for one episode, \abbr{ie}, for one optimization run. The random expansion step enhances the ability of the fully trained agent to generalize, by demanding it to deal well with a broad spectrum of both inefficient and efficient circuits, which will be encountered during every successful optimization run.

\paragraph{Simulated annealing} As a benchmark for the results by the RL agent, we consider simulated annealing as an alternative approach, using an adapted Metropolis rule and an exponential cooling schedule. We apply simulated annealing to the randomly expanded circuits, \abbr{ie}, on the same dataset as later for the agent. The results depend strongly on the number of steps: Within $10000$ steps, simulated annealing is able to merely reach the level after pruning (on average). Within $200000$ steps, out of which roughly $15\hdots\SI{20}{\percent}$ are actually accepted, a mean depth $\langle d\rangle=27.35\pm0.08$ and gate count $\langle n\rangle=105.15\pm0.32$ can be achieved.

\paragraph{RL} Now, we apply the RL approach as described in \cref{sec:technique} on our dataset of expanded random circuits. The hyperparameters have been optimized in a systematic search. One full learning process takes $6$ to $7$ days, on a node with $32$ CPU cores.

We find that the learning process can be divided roughly into two phases (\abbr{cmp} \cref{fig:results_small_rand_circs}c): In the first \abbr{ca} $50$ epochs, the agent rapidly improves in minimizing the mean depth and gate count, to a level of $\langle d\rangle\approx29$ and $\langle n\rangle\approx100$ which is already considerably better than after pruning. In the remaining course of the training, the agent can slowly, but steadily improve towards $\langle d\rangle=27.20\pm0.07$ and $\langle n\rangle\approx97.86\pm0.33$ at around epoch $1000$. Hence, the agent can clearly beat the results from simulated annealing, especially regarding the gate count. Because we use a new circuit in each training episode, no difference is to be expected between the performance on training and validation samples.

In \cref{fig:results_small_rand_circs}d, we examine closer how a trained agent proceeds in optimizing a circuit. For this purpose, we choose $3200$ circuits and plot how $d$ and $n$ evolve with every transformation. As \cref{fig:results_small_rand_circs}d shows, the agent can rapidly improve the circuits during the first $150$ to $200$ steps. Afterwards, the curve is mostly flat, except for two effects: First, there are occasional excursions into lower-quality states, but the agent usually returns quickly to the previous level. Second, the agent sometimes manages to further optimize the circuit after some time, partially via passing some intermediate lower-quality states, \abbr{ie}, to cross a reward barrier. Note that we did not give the agent the ability to terminate an episode.

Compared to the pruning level, we have seen that the agent achieves much better values for $\langle d\rangle$ and $\langle n\rangle$ on average. How effective the agent has actually become can be seen by comparing $d$ and $n$ for individual circuits between the agent and after pruning. In \cref{fig:results_small_rand_circs}e, we plot this comparison for epochs $900-1000$, \abbr{ie}, for $3200$ circuits in total. The agent achieves a reduction of both $d$ and $n$ for the vast majority of circuits, partially up to even more than $\SI{50}{\percent}$. Only for $12$ out of $3200$ circuits, \abbr{ie}, in around $\SI{0.4}{\percent}$ of the cases, the agent fails to achieve a better $q(s)$ than after pruning.

\subsection{Extrapolation to larger random circuits}
\label{sec:results:extrapolation_to_large}

\begin{figure}
	\includegraphics[width=\columnwidth]{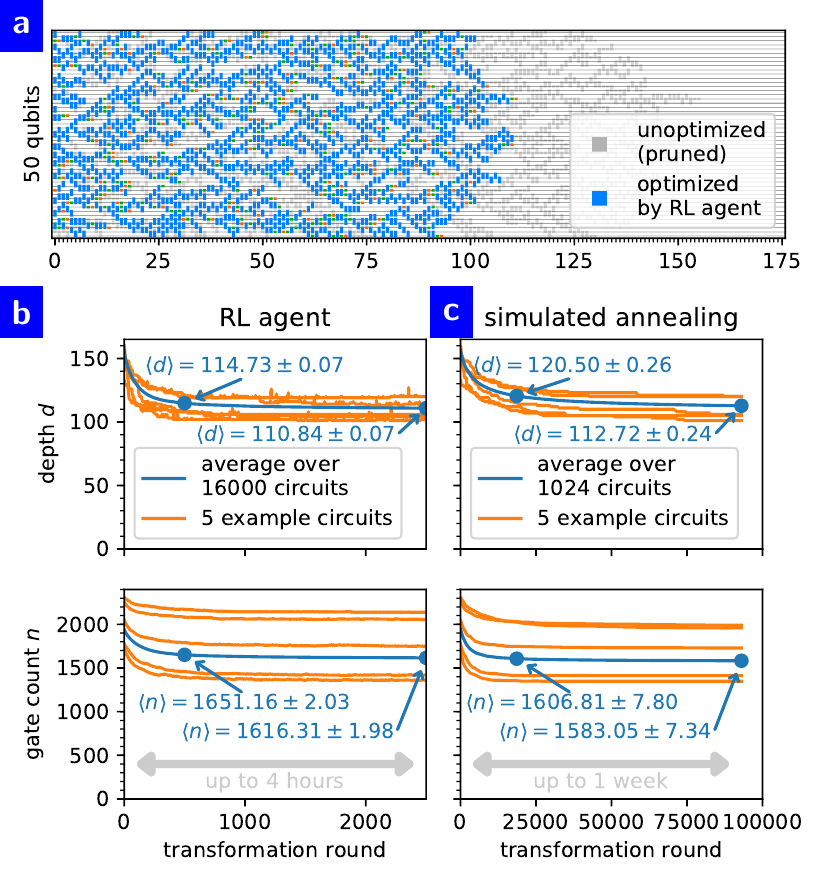}
	\caption{
		Extrapolation to $50$-qubit random circuits. The agent has been trained on $12$-qubit circuits (\abbr{cmp} \cref{fig:results_small_rand_circs}), no further learning updates are performed here.
		(a) shows the comparison between an unoptimized example circuit (after pruning) and the result of the optimization by the RL agent.
		(b) shows the progress of the agent in reducing depth $d$ and gate count $n$ over the course of $2500$ transformations.
		(c) shows the corresponding curves for simulated annealing, which requires almost $100000$ transformations to achieve a comparable degree of optimization (the computation was terminated after 1 week, at transformation $93000$).
	}
	\label{fig:results_large_rand_circs}
\end{figure}

In \cref{sec:results:rand_circs}, the agent has been trained and evaluated on $12$-qubit random circuits. However, due to its fully convolutional architecture, this very same agent can -- at least in principle -- be directly used to optimize also circuits of a different size, in particular much larger ones. Now, we will test how well the agent can actually generalize its knowledge in this situation.

For this purpose, we reuse the scheme to generate random circuits as described in \cref{sec:results:rand_circs}, except for changing two parameters: we increase the number of qubits from $12$ to $50$, and the number of initial gates from $150$ to $2500$. We find $\langle d\rangle=199.25\pm0.08$ and $\langle n\rangle=2655.3\pm1.2$ before pruning, and $\langle d\rangle=156.67\pm0.07$ and $\langle n\rangle=1940.2\pm1.6$ after pruning. Because we will not use the circuits here to train the agent, we can skip the step to expand them by random transformations, whose purpose it was to feed to the agent also very inefficient circuits during training. Instead, the optimization by the agent starts here directly from the pruned circuits. As shown in \cref{fig:results_large_rand_circs}b, the agent achieves to reduce $\langle d\rangle$ to $110.84\pm0.07$ and $\langle n\rangle$ to $1616.3\pm2.0$ within $2500$ transformations. Remarkably, the reduction ratio in these two quantities is comparable to the smaller circuits it has been trained on (\abbr{cmp} \cref{fig:results_small_rand_circs}).

Simulated annealing arrives at similar values, $\langle d\rangle=112.72\pm0.24$ and $\langle n\rangle=1583.0\pm7.3$, within $93000$ transformations. These are much fewer transformations than required to optimize the smaller random circuits in \cref{sec:results:rand_circs}, probably because here the random expansion step has been skipped. Nevertheless, $93000$ transformations for each larger random circuits here have already taken one week (our termination criterion), which is comparable to the time needed to train an RL agent. Afterwards, this agent can optimize arbitrary circuits, in a relatively short time ($3\hdots\SI{5}{\hour}$ in this case).

Our results show that an agent can actually extrapolate its knowledge to larger circuits. More generally, they demonstrate that our approach, both with RL and simulated annealing, works deep in the quantum supremacy regime. Furthermore, this also highlights a situation where optimizing even a single circuit with simulated annealing needs already a runtime comparable to the full training of an RL agent and subsequently optimizing the particular circuit.

\subsection{QAOA-MaxCut circuit}
\label{sec:results:qaoa_maxcut}

\begin{figure}
	\includegraphics{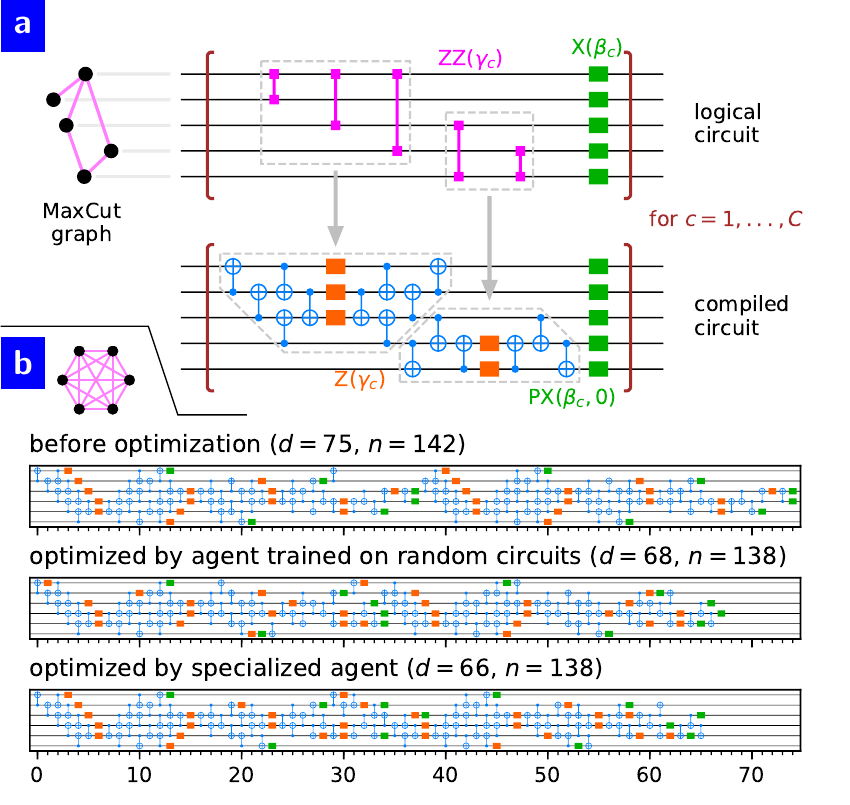}
	\caption{
		Optimization of QAOA-MaxCut circuits.
		(a) indicates how to translate the MaxCut problem for a graph into a quantum circuit following QAOA, and how to efficiently compile this logical circuit into our gate set. We display one of $M$ cycles which form the full circuit, each with a different set of parameters $(\gamma_c,\beta_c)$ whose values are refined during the QAOA algorithm.
		(b) shows the compiled circuit for $C=2$ cycles and an all-to-all-connected graph with $6$ nodes, which has depth $d=75$ and gate count $n=142$ (top). Using a generic agent trained on random circuits as in \cref{fig:results_small_rand_circs}, we find (by postselection) improved circuits with $d=68$ and $n=138$ (middle). A specialized agent trained on this particular circuit can further optimize it to $d=66$ and $n=138$ (bottom).
	}
	\label{fig:results_qaoa_maxcut}
\end{figure}

As an example for a real-world quantum algorithm, we now consider the MaxCut problem. The goal is to arrange the nodes of an undirected, non-weighted graph into two groups such that the amount of cut edges is maximized. Finding the exact solution is an NP-hard problem. Following the quantum approximate optimization algorithm (QAOA \cite{farhi2014qaoa}), approximate solutions can be found with the help of a quantum circuit consisting of repeated cycles of ZZ gates and local X rotations with variable angles \cite{harrigan2021qaoa} (\abbr{cmp} \cref{fig:results_qaoa_maxcut}a). We consider the same gate set as in the examples above, such that we can reuse the previously trained agent. Also, this covers the realistic situation where the native gates of the quantum algorithm do not necessarily match the native gates of the hardware. \Cref{fig:results_qaoa_maxcut}a shows an efficient compilation of this circuit into our gate set, where ZZ gates need to be decomposed into CNOTs and local Z rotations (local X rotations are a special case of Phased-X gates). Note that the variable angles of the gates do not affect the optimization strategy, as long as we assume these angles to be generic (\abbr{ie}, not set to special values which would allow additional optimizations).

Every graph corresponds to a different circuit. We restrict our analysis here to all-to-all connected graphs. While any such graph has trivial MaxCut solutions (nodes evenly distributed), the corresponding circuit is still useful for us to consider because the circuit for any other graph layout can be derived from it by removing some of the ZZ gates; therefore, it also provides a worst-case estimate. In \cref{fig:results_qaoa_maxcut}b, we specifically consider a graph with $6$ nodes and $2$ cycles in the QAOA circuit.

First, we try to optimize this QAOA circuit with a generic agent trained on $12$-qubit \textit{random} circuits, as in \cref{fig:results_small_rand_circs}. We find that on average, the agent manages to slightly reduce the gate count, although the depth is slightly increased. Even though this agent is, thus, not able to reliably optimize the circuit, it is still able to sometimes achieve an improvement: By postselecting over all time steps in multiple optimization runs, we find that the best circuit reduces the depth from $75$ to $68$ and, simultaneously, the gate count from $142$ to $138$ (\abbr{cmp} \cref{fig:results_qaoa_maxcut}b); note that different runs deviate because the policy is to some degree stochastic, and that the circuit quality fluctuates also with time. We observe that the optimizations happen at the interface between subsequent cycles, and at the end of the last cycle.

If we, however, train an agent via RL directly on this specific circuit, it does not only learn to reliably improve the circuit, but it also finds two further optimizations, each reducing the depth by $1$. These optimizations are very hard to discover, as they require a lookahead by $7$ and $9$ transformations, respectively, where the search breadth is around $350$. The resulting circuit, as shown in \cref{fig:results_qaoa_maxcut}b, has a depth of $66$ and a gate count of $138$. To our knowledge, this is the most efficient realization on the gate set we provided (when including additional gates, one can arrive at an even shorter circuit \cite{harrigan2021qaoa}).

For these QAOA-MaxCut circuits, we start from an already quite efficient compilation, so the optimization potential is not very large here. Nevertheless, the agent is still able to find optimizations, including some which require complex transformation sequences and would thus be very hard to find by hand. On the general level, this example demonstrates that our approach can also be used to optimize real-world circuits, where the optimization potential might very well be larger than here. However, we have got reliable improvement and the best optimization quality only for agents trained on this circuit, which emphasizes the importance of a proper training dataset.

During early simulations, we made a noteworthy observation: Sometimes, the agent trained on these MaxCut circuits achieved optimizations far below a depth of $66$ and a gate count of $138$. It turned out that this happened only if the rotation angle for one cycle of ZZ gates, which was chosen at random, was relatively close to a multiple of $\mconst{pi}$. In this case, the agent found a creative way to exploit that the closeness criterion for one of the transformation rules was too loose, which it used to partially or completely remove this cycle from the circuit. Although in this particular case these transformations were not valid, the agent's behavior shows that it would readily exploit the simplification potential in other complex circuits.

\subsection{Discussion}
\label{label:results:discussion}

\paragraph{Dataset} We anticipate our RL agents to be considerably more powerful and reliable on real-world quantum circuits when trained on a dataset with similar properties, which is also indicated by our results in \cref{sec:results:qaoa_maxcut}. In contrast to currently existing collections of quantum circuits, like \cite{url_pyzx_dataset}, we need for our purposes first the circuits being compiled for specific hardware architectures, and second, also a significantly larger dataset size to fully train agents on them (based on \cref{sec:results:rand_circs}, this can require in the order of $10000$ to $100000$ episodes).

We think of our future dataset to be generated as follows. We consider a set of various quantum algorithms, like QAOA \cite{farhi2014qaoa}, variational quantum eigensolver \cite{mcclean2016vqe}, Shor's algorithm \cite{shor1999factorization_algo} (for small integers), and many more \cite{url_qu_algo_zoo}. Many of these algorithms imply a large number of different circuits; for example, QAOA can be applied to multiple combinatorial optimization problems, and for each of them, the circuit changes with the concrete problem instance (\abbr{eg}, graph layout) and the number of cycles. These circuits are described in a high-level, hardware-independent way. For every specific architecture (gate set, layout, connectivity), we would then generate an own dataset by compiling the circuits for them, which can be done using existing quantum libraries like Cirq \cite{url_cirq}.

Eventually, each of our agents will be trained on one of these architecture-specific datasets. If the dataset size is not yet sufficient for this, we can use data augmentation techniques (like mirroring the circuit in space and/or time), and to some degree also repeat circuits during training.

\paragraph{RL \abbr{vs} simulated annealing} In \cref{sec:results:rand_circs}, simulated annealing achieved $98.48\pm1.06\%$ of the reduction in gate count $n$ by the RL agent, but only $58.06\pm2.15\%$ in the depth $d$; \cref{sec:results:extrapolation_to_large} is more ambiguous, as simulated annealing achieved $109.67\pm2.44\%$ of the reduction in $n$ by the RL agent, but only $95.88\pm0.57\%$ in $d$ (all numbers in reference to the pruning level). Therefore, in these scenarios, RL tends to give overall slightly better results than simulated annealing.

Another aspect to consider in this comparison is runtime. After having been trained once on a certain class of circuits, an agent is considerably faster than simulated annealing in optimizing arbitrary circuits of this class (\abbr{ca} $\SI{2}{\minute}$ \abbr{vs} $1\hdots\SI{3}{\day}$ in \cref{sec:results:rand_circs}, $3\hdots\SI{5}{\hour}$ \abbr{vs} $\SI{7}{\day}$ in \cref{sec:results:extrapolation_to_large}). Of course, for a fair comparison of the computational effort, it is necessary to also take into account the time needed for training the agent in the first place (up to $1$ week), and eventually also the time to optimize the hyperparameters for both approaches. Therefore, if we take all those aspects into account, the question which method is less computationally expensive cannot be answered in general, but depends on the class of circuits and previously solved problem instances -- the only safe statement is that for a sufficient number of circuits to be optimized, the RL approach will be computationally cheaper.

\paragraph{Symbolic parameters} In our current implementation, gates are represented numerically, which requires to consider fixed choices for the gate parameters. This is not directly a property of the RL approach presented in this work, but of our underlying quantum circuit framework \cite{url_rl4circopt}. With manageable effort, this framework can be extended to allow also gates parameterized by symbolic variables. This would simplify the optimization of entire classes of parameterized circuits, like for QAOA with its varying angles $\beta_c$ and $\gamma_c$. So far, this is only indirectly possible by repeating the same sequence of transformations for differently parameterized circuits.

\section{Conclusion}

In this article, we have introduced RL as a powerful and flexible approach to QCO. In particular, we have focused on the optimization of the local circuit structure to the available hardware resources, which is especially relevant for the NISQ era. The main challenge here is to discover suitable sequences of circuit transformations; the size of the search space and the lack of simple strategies in many situations, especially for circuits with an irregular structure, make this a difficult optimization problem. RL is well-suited to problems of this kind, and promises to simultaneously fulfill the following key features: \textit{hardware-efficient}, \abbr{ie}, that it can find the optimal solution given the available resources; \textit{cross-platform}, \abbr{ie}, that the approach works for different quantum architectures (gate sets, chip layouts, connectivities, \abbr{etc}), without large migration costs; \textit{autonomous}, \abbr{ie}, that the only human effort is to specify the problem, and the optimization is done by the computer completely on its own, without requiring further interaction; and \textit{reliable}, \abbr{ie}, that the results are always correct and for a wide spectrum of circuits (close-to-)optimum.

For our approach to be successful, we had to address several key challenges: First, we have formulated QCO as a problem which can be efficiently addressed using RL (\abbr{cmp} \cref{sec:technique:qco_as_rl_problem,sec:technique:circ_trafos}). Second, we have developed efficient schemes to represent circuits for the network input and transformations for the network output (\abbr{cmp} \cref{sec:technique:network_io}). Third, as prerequisite for applying techniques like RL or simulated annealing, we have implemented a Python framework to automate the process of identifying all transformations within a circuit \cite{url_rl4circopt}.

Both RL and simulated annealing turn out to be valid optimization techniques for the QCO problems considered in this work. In terms of optimization quality, the difference between both approaches in reducing depth and gate count were not larger than $10\%$ (in reference to the pruning level), except for \cref{sec:results:rand_circs} where simulated annealing achieved only $58.06\pm2.15\%$ of the depth reduction by the RL agent (\abbr{cmp} \cref{label:results:discussion}). In terms of computational effort, our RL agents required up to $1$ week of training time, but were afterwards considerably faster than simulated annealing in optimizing arbitrary circuits from the class they have been trained on (\abbr{ca} $\SI{2}{\minute}$ \abbr{vs} $1\hdots\SI{3}{\day}$ in \cref{sec:results:rand_circs}, $3\hdots\SI{5}{\hour}$ \abbr{vs} $\SI{7}{\day}$ in \cref{sec:results:extrapolation_to_large}). We expect that RL can deal better with situations where decision-making becomes even harder than here. Therefore, RL seems to be, especially in the long run, the more promising approach.

So far, we have mostly considered randomly generated circuits to circumvent the lack of a sufficient dataset of realistic circuits; one central future direction will be to collect such a dataset. Several direct extensions of our work are straightforward to implement, yet potentially very fruitful for future applications, such as implementations for different gate sets, 2D geometries, \abbr{etc} A (semi-)automated procedure to identify transformation rules, where machine learning could be beneficial, would further reduce the effort for including new gate sets. The optimization of parameterized circuit classes can be simplified by implementing symbolic variables as gate parameters in the underlying quantum circuit framework. Furthermore, the existing complementary QCO techniques which focus on optimizing the global circuit structure \cite{amy2014t_par,amy2019t_par,heyfron2018t_opt,zhang2019t_optimizer,kissinger2020py_zx} could be combined in a two-stage process with our technique to optimize the local circuit structure.

We expect our approach to soon become a standard tool for optimizing circuits in NISQ experiments. Central practical advantages are the negligible additional effort to consider non-ideal circumstances, like inhomogeneous error rates or defect qubits, and the ability to customize the optimization criterion.

\paragraph{Data and code availability} The code is publicly available under \cite{url_rl4circopt}. Other findings of this study are available from the corresponding author on reasonable request.

\paragraph{Acknowledgements} T.\,F.\ thanks Google Research for hosting during his internship.

\paragraph{Competing interests} The authors declare no competing interests.

\paragraph{Corresponding author} Correspondence should be addressed to T.\,F.

\bibliography{references}{}
\bibliographystyle{unsrt}

\end{document}